\documentclass[a4paper,10pt]{article}
\usepackage{graphicx} 
\usepackage{graphics}

\usepackage{amsmath}   
\usepackage{amsfonts}
\usepackage{amssymb}

\newcommand{\dfr}{\bar{d}} 

\begin{document}
%
%
\def\ov{\over}
\def\l{\left}
\def\r{\right}
\def\be{\begin{equation}}
\def\ee{\end{equation}}
%

\setlength{\oddsidemargin} {1cm}
\setlength{\textwidth}{18cm}
\setlength{\textheight}{23cm}
\title{The Pauli equation in scale relativity}
\author{{\bf Marie-No\"elle C\'el\'erier$^1$ and Laurent Nottale$^2$} 
\\
LUTH, CNRS, Observatoire de Paris-Meudon, \\
5 place Jules Janssen, 92195 Meudon Cedex, France \\
e-mail: $^1$ marie-noelle.celerier@obspm.fr \\
$^2$ laurent.nottale@obspm.fr}


\maketitle

\begin{abstract}

In standard quantum mechanics, it is not possible to directly extend the Schr\"odinger equation to spinors, so the Pauli equation must be derived from the Dirac equation by taking its non-relativistic limit. Hence, it predicts the existence of an intrinsic magnetic moment for the electron and gives its correct value. In the scale relativity framework, the Schr\"odinger, Klein-Gordon and Dirac equations have been derived from first principles as geodesics equations of a non-differentiable and continuous spacetime. Since such a generalized geometry implies the occurence of new discrete symmetry breakings, this has led us to write Dirac bi-spinors in the form of bi-quaternions (complex quaternions). In the present work, we show that, in scale relativity also, the correct Pauli equation can only be obtained from a non-relativistic limit of the relativistic geodesics equation (which, after integration, becomes the Dirac equation) and not from the non-relativistic formalism (that involves symmetry breakings in a fractal 3-space). The same degeneracy procedure, when it is applied to the bi-quaternionic 4-velocity used to derive the Dirac equation, naturally yields a Pauli-type quaternionic 3-velocity. It therefore corroborates the relevance of the scale relativity approach for the building from first principles of the quantum postulates and of the quantum tools. This also reinforces the relativistic and fundamentally quantum nature of spin, which we attribute in scale relativity to the non-differentiability of the quantum spacetime geometry (and not only of the quantum space). We conclude by performing numerical simulations of spinor geodesics, that allow one to gain a physical geometric picture of the nature of spin.

\end{abstract}




\section{Introduction}
\label{s.i}


The theory of scale relativity, when applied to microphysics, allows one to recover quantum mechanics as a non-classical mechanics on a non-differentiable spacetime. In this framework, the Schr\"odinger, Klein-Gordon and Dirac equations have been derived in terms of integrals of geodesics equations \cite{LN93,LN96,CN03,CN04}. The complex nature of the wavefunction in the Schr\"odinger and Klein-Gordon equations proceeds from the necessity to introduce, because of the non-differentiability, a discrete symmetry breaking on the (proper) time differential element. The bi-quaternionic nature of the Dirac bi-spinors arises from further symmetry breakings on the spacetime variables, which also proceed from non-differentiability.

Now, it is well known that the properties of the non-relativistic 
limit of a dynamical equation may differ from those obtained 
when the limiting equation is based directly on exact Galilean 
kinematics, see e.g. \cite{HB03}. This is in particular the case for the Pauli equation which predicts the existence of an intrinsic 
magnetic moment for the electron and gives its correct value only 
when it is obtained as the non-relativistic limit of the Dirac 
equation.

We therefore wish to test if this property is also valid in the 
framework of scale relativity. In this theory, the dynamical 
equations of quantum mechanics proceed from successive discrete symmetry breakings on the differential elements (which are a consequence of the non-differentiable geometry) and on the spacetime variables. The Schr\"odinger equation is obtained from the breaking of the symmetry $dt \leftrightarrow -dt$ \cite{LN93,LN96}, the Klein-Gordon one from the symmetry breaking $ds \leftrightarrow -ds$ \cite{LN96}, and the Dirac equation from the symmetry breakings $ds \leftrightarrow -ds$, $dx^{\mu} \leftrightarrow -dx^{\mu}$ and $x^{\mu} \leftrightarrow -x^{\mu}$ \cite{CN03,CN04}. Therefore, one could be led to ask whether the Pauli equation might proceed from the symmetry breakings $dt \leftrightarrow -dt$ (for a non-relativistic motion) and $dx^{\mu} \leftrightarrow -dx^{\mu}$ (for the appearence of spinors).

After having derived in section~\ref{s.scn} the existence of spinors as a geometric and algebraic consequence of the giving up of the spacetime differentiability hypothesis, we answer in section~\ref{s.psb} to this question and show that, as in standard quantum mechanics, the Pauli equation cannot be obtained from a non-relativistic symmetry breaking procedure. We therefore give, in section~\ref{s.ped}, the proper way to derive the Pauli equation from the Dirac equation, by applying to the scale relativity tools the standard method of quantum mechanics. In section~\ref{s.srp}, we propose a new mathematical representation of the scale relativistic tools needed to build the Dirac equation and spinor, the relevance of which is based upon the results of its degeneracy towards the Pauli ones. We obtain, as a consequence, the Pauli scale relativistic ingredients as mere non-relativistic limits of the Dirac ones. Finally, we present in section~\ref{s.gfs} the results of numerical simulations of geodesics of a fractal spacetime which carry an intrinsic angular momentum and discuss the physical picture of the nature of spin that therefore emerges in the scale relativity framework. Section \ref{s.dac} is devoted to the discussion and conclusion.


\section{Spinors as a consequence of non-differentiability}
\label{s.scn}


In the theory of scale relativity, the motion equations of quantum mechanics are derived as geodesics equations in a non-differentiable and continous space(time) (which can subsequently be proved to be fractal). 

The non-motion-relativistic Schr\"odinger equation is obtained from the giving up of the coordinate differentiability hypothesis in a fractal 3-space, where the time $t$ is the Galilean invariant time. Non-differentiability implies the breaking of the symmetry $dt \leftrightarrow -dt$ \cite{LN93,LN96}. This breaking leads us to use complex numbers for a covariant representation of quantum variables, and, in particular, of the wavefunction \cite{CN04}.

The motion-relativistic Klein-Gordon equation proceeds from the 
giving up of the coordinate differentiability in a fractal 4-spacetime with the line element $s$ as a curvilinear parameter. It implies the breaking of the symmetry $ds \leftrightarrow -ds$ \cite{LN96}. This breaking has the same consequence as regards the use of complex numbers for a covariant representation of quantum variables, and, in particular, of the wavefunction \cite{CN04}.

We have also obtained the motion-relativistic Dirac equation by giving up the coordinate differentiability in a fractal 4-spacetime with the line element $s$ as a curvilinear parameter. It proceeds from 
the breaking of further symmetries besides $ds \leftrightarrow 
-ds$, i.e., $dx^{\mu} \leftrightarrow -dx^{\mu}$ and $x^{\mu} 
\leftrightarrow - x^{\mu}$ \cite{CN03,CN04}. These successive 
symmetry breakings involve algebra doublings that led us to use 
bi-quaternions to describe bi-spinors and the Dirac equation 
\cite{CN04}.

Now, it is well known that on manifolds exhibiting a quaternionic 
structure, spinors appear automatically \cite{SA95}. Because the quaternion algebra is isomorphic to the rotation group algebra 
$SO(3)$, itself related to $SU(2)$ by homomorphism, we can consider the two one-dimensional irreductible representations of $SU(2)$ over the quaternions. One is the trivial spin-0 representation. The 
other is the spin-1/2 representation which acts on a one-component 
quaternion basis $\phi$. 

Instead of writing this quaternion as a function of its four 
components: $\phi = \phi_0 + i \,  \phi_1 + j \,  \phi_2 + k \,  \phi_3$, we can write it as a function of its symplectic components. These are elements of a complex subspace of the quaternionic algebra 
spanned by $1$ and $i$ and denoted by $\mathbb{C}(1,i)$. In the symplectic representation $\phi$ is written as $\phi = \phi_{\alpha} + j \,  \phi_{\beta}$, the symplectic components being defined as $\phi_{\alpha} = \phi_0 + i \,  \phi_1$ and $\phi_{\beta} = \phi_2 - i \,  \phi_3$.

The induced action of the generators of the spin-1/2 representation 
of $SU(2)$ on the two-component column vector
\begin{displaymath}
\Phi = \left (
\begin{array}{c}
\phi_{\alpha} \\
\phi_{\beta}
\end{array}
\right ), 
\end{displaymath}
is given by the $2 \times 2$ Pauli spin matrices. In other words, the 
non-trivial one-dimensional quaternionic irreducible representation 
of $SU(2)$ induces the appearance of two-component complex spinorial representations. Therefore, the apparent structural doubling associated with complex spinors arises automatically in the framework of manifolds carrying a quaternionic structure.

In scale relativity, both the complex nature of the wavefunction and the appearance of spin have a common origin, namely the fundamental two-valuedness of the derivatives issued from non-differentiability. However, while the origin of the complex nature of the wavefunction is linked to non-differentiability at the level of the total derivative, the origin of spin is due to non-differentiability at the level of partial derivatives with respect to the coordinates. These two successive doublings are naturally accounted for in terms of algebra doublings \cite{CN04}.

To illustrate this statement, we consider the total derivative with respect to the time $t$ of a differentiable function $f(t)$. It can be written twofold:
\begin{equation}
\frac{df}{dt} = \lim_{dt \rightarrow 0}\frac{f(t+dt) - f(t)}{dt} = 
\lim_{dt \rightarrow 0}\frac{f(t) - f(t-dt)}{dt} \; .
\label{eq.1}
\end{equation}
The two definitions are equivalent in the differentiable case. In the non-differentiable situation, both definitions fail, since the limits are no longer defined. In the framework of scale relativity, the physics is related to the behaviour of the function during the `zoom' operation on the time resolution $\delta t$, which is now considered as an independent variable. We therefore replace $f(t)$ by a fractal function $f(t,\delta t)$, explicitly dependent on the time resolution interval, whose derivative is undefined only at the unobservable limit $\delta t \rightarrow 0$. As a consequence of the very construction of the derivative, which needs two points to be defined (instead of one for the position coordinates), there are two definitions of the derivative of a fractal function instead of one.  In the theoretical description considered here (which may be different from an experimental siuation), the resolution  $\delta t$ can be identified with the differential element $dt$ \cite{CN04}. Therefore two functions $f'_+$ and $f'_-$ are introduced as explicit functions of the two variables $t$ and $dt$
\begin{equation}
f'_+(t,dt) = \frac{f(t+dt,dt)-f(t,dt)}{dt} \; ,
\label{eq.2}
\end{equation}
\begin{equation}
f'_-(t,dt) = \frac{f(t,dt)-f(t-dt,dt)}{dt} \; .
\label{eq.3}
\end{equation}
We pass from one to the other by the transformation $dt \leftrightarrow -dt$ (local differential time reflection invariance), which is an implicit discrete symmetry of differentiable physics. Namely, the transformation $dt \leftrightarrow -dt$ is nothing but  the discrete version of the continuous scale transformation $dt \rightarrow dt'$ which plays the central role in the scale relativity theory. The non-differentiable geometry of scale relativity now implies that this symmetry is broken, which corresponds to the doubling $d/dt \rightarrow (d_+/dt, d_-/dt)$. To recover local differential time reversibility in terms of a new complex 
process, we combine the two derivatives in terms of a complex derivative operator. This is, in this framework, the origin of the complex nature of the wavefunction of quantum mechanics \cite{LN93,CN04}.

To go further on in the construction of the theory, we are led to consider that the velocity fields of the geodesics bundles, which 
intervene in the definition of the wavefunction, are functions of the coordinates. Therefore, we need to also analyse the physical meaning of the partial derivatives $\partial/\partial x$ (we use only one coordinate variable in order to simplify the writing, but this applies to all coordinates) in the decomposition $d/dt=\partial/\partial t+(dx/dt)\,  \partial/\partial x$. Strictly speaking, $\partial f / \partial x$ does not exist in the non-differentiable case. We are therefore once again led to introduce fractal functions $f(x,\delta x)$, explicitly dependent on the coordinate resolution interval, whose derivative is undefined only at the unobservable limit $\delta x \rightarrow 0$. As for the case of the total derivative, we identify $\delta x$ with the differential element $dx$ and consider the two definitions of the partial derivative of a fractal function, namely
\begin{equation}
{\frac{\partial f }{\partial x}}_{+} = {\frac{f(x+dx,dx)-f(x,dx)}{dx}},
\label{eq.4}
\end{equation}
\begin{equation}
{\frac{\partial f}{\partial x}}_{-} = {\frac{f(x,dx)-f(x-dx,dx)}{dx}}.
\label{eq.5}
\end{equation}
They are transformed one into the other under the reflection $dx \leftrightarrow -dx$, which is also an implicit discrete symmetry of differentiable physics. The non-differentiable geometry of scale relativity implies that this symmetry is broken which corresponds to the new doubling $\partial/\partial x \rightarrow (\partial_+/\partial x, \partial_-/\partial x)$. This finally leads to a doubling of the wavefunction itself $\psi \rightarrow (\psi_1, \psi_2)$ characterizing a Pauli spinor. We have already mentioned that these two successive doublings can be naturally accounted for in terms of algebra doublings, i.e., a description tool that jumps from real numbers $\mathbb{R}$ to complex numbers $\mathbb{C}=\mathbb{R}^2$ then to quaternions $\mathbb{H}=\mathbb{C}^2$. Now, as recalled above, spinors appear automatically on manifolds exhibiting a quaternionic 
structure. Spinors are therefore natural structures implied by the geometric and algebraic consequences of non-differentiability.


\section{Can the Pauli equation be obtained from symmetry breakings in the framework of scale relativity?}
\label{s.psb}


Now, the question is: does the Pauli equation naturally arise from the standard scale relativistic construction of a non-relativistic motion equation as a geodesics equation involving two successive doublings in the framework of a fractal 3-space?

The relevant formalism will be that developed for the derivation of the non-motion-relativistic equations of quantum mechanics, valid in a fractal 3-space with the time $t$ as a curvilinear parameter \cite{LN96}.


\subsection{Transition from non-differentiability to  differentiability}


When we apply the reasoning of section~\ref{s.scn} to the 3-space coordinates, generically denoted by $X$, we see that the velocity 
\begin{equation}
V = {\frac{dX}{dt}}= \lim_{dt \rightarrow 0} \frac{X(t+dt) - X(t)}{dt}
\label{eq.6}
\end{equation}
is undefined. But it can be redefined in a new way as a fractal function $V(t,dt)$. The scale dependence of the velocity suggests that we complete the standard equations of physics by new differential equations of scale. Writing the simplest possible equation for the variation of the velocity in terms of the scale variable $dt$, as a first order differential equation $dV/d \ln dt=\beta(V)$, then Taylor expanding it, using the fact that $V<1$ (in motion-relativistic units $c=1$), we obtain the solution as a sum of two terms: a scale-independent, differentiable, `classical' part and a power-law divergent, explicitly scale-dependent, non-differentiable `fractal' part \cite{LN97A},
\begin{equation}
V = v + w = v \; \left[1 + a \left(\frac{\tau}{dt}\right)^{1-1/D_F}\right],
\label{eq.7}
\end{equation}
where $D_F$ is the fractal dimension of the path.

The transition scale $\tau$ yields two distinct behaviours of the velocity depending on the resolution at which it is considered, since $V \approx v$ when $dt \gg \tau$ and $V \approx w$ when $dt \ll \tau$. In the following case when this description holds for a quantum particle of mass $m$, $\tau$ is identified with the de Broglie scale of the system ($\tau=\hbar/E$) and the explicit `fractal' domain with the quantum one. But it should be emphasized that, in our description, the geometry is actually fractal at all scales, even though the fractal contribution $w$ becomes dominated by the classical one $v$ at scales larger than the de Broglie scale: this is an important point (which may have been unclear in \cite{CN04}), since it may lead to specific tests of the theory (e.g., by searching for very faint residual quantum contributions in the classical domain). The scaled fluctuation $a$ is described by a dimensionless stochastic variable which is normalized according to $\langle a \rangle = 0$ and $\langle a^2\rangle = 1$. Recalling that $D_F = 2$ plays the role of a critical dimension \cite{LN93,LN96}, we shall here consider only the case of this fractal dimension $2$.

The above description applies to any of the fractal geodesics. Now, one of the geometric consequences of the fractal character of space is that there is an infinity of fractal geodesics relating any couple of its points \cite{LN93,LN96}. It has therefore been suggested \cite{LN89} that the description of a quantum mechanical particle could be reduced to the geometric properties of the set of fractal geodesics that corresponds to a given state of this `particle'. As a consequence, any measurement is interpreted as a sorting out (or selection) of the geodesics bundle due to the interaction with the measuring device \cite{LN93,LN89} and/or linked to the information known about the system.

Equation (\ref{eq.7}) multiplied by $dt$ gives the elementary displacement, $dX$, of the system as a sum of two terms,
\begin{equation}
dX = dx + d\xi,
\label{eq.8}
\end{equation}
$d\xi$ representing the `fractal part' and $dx$, the `classical part', defined, for $D_F=2$, as
\begin{equation}
dx = v \; dt, 
\label{eq.9}
\end{equation}
\begin{equation}
d\xi=a \, \sqrt{2 {\cal D} \, dt},
\label{eq.10}
\end{equation}
with $2{\cal D}=\tau v^2$. Owing to equation~(\ref{eq.7}), we identify $\tau$ as the (non-relativistic) Einstein transition scale, $\hbar / E = \hbar / {\frac{1}{2}} mv^2$, and therefore $2{\cal D}$ can be identified with the Compton scale (for $c=1$), so that  ${\cal D}=\hbar / 2m$. We note, from equations~(\ref{eq.8})-(\ref{eq.10}), that $dx$ scales as $dt$, while $d\xi$ scales as $dt^{\frac{1}{2}}$. The elementary displacement on a fractal space is therefore, as expected, the sum of a `classical', differentiable element, $dx$, which is leading at large scales, and a `fractal', diverging fluctuation, $d\xi$, which is leading at small scales.


\subsection{Fractal velocity fields and symmetry breaking $dt \leftrightarrow -dt$}


In order to account for the infinite number of geodesics, the velocity should now  be defined as a fractal velocity field. Moreover, as recalled in section~\ref{s.scn}, non-differentiability implies a two-valuedness of this velocity field, that we describe by fractal functions of space coordinates and time, i.e., explicit functions of the resolution interval $dt$, namely $V_+[x(t,dt),t,dt]$ and $V_-[x(t,dt),t,dt]$. These two velocity fields can be in turn decomposed in terms of a `classical part', which is differentiable and independent of resolution, and a `fractal part', $V_{\pm}[x(t,dt),t,dt]=v_{\pm}[x(t),t]+w_{\pm}[x(t,dt),t,dt]$. Note that there is no reason {\it a priori} for the two classical velocity fields themselves to be equal. 

Recall that this two-valuedness of the velocity vector finds its origin in a breaking of the discrete time differential element reflection invariance symmetry ($dt \leftrightarrow -dt$), which is itself a mathematical consequence of non-differentiability. If one reverses the sign of this time differential element, $v _{+}$ becomes $v _{- }$.


\subsection{Symmetry breakings $dx \leftrightarrow -dx$ and $dt \leftrightarrow -dt$}


The three minimal effects of non-differentiability considered above lead to the construction of a complex wavefunction that is solution of a Schr\"odinger equation \cite{LN93,CN04}. But a more general description should also include another consequence of non-differentiability, namely the discrete symmetry breaking of the reflection $dx \leftrightarrow -dx$ on the space differential elements. Let us now consider this case, which leads to introduce a quaternionic representation of physical quantities.

In the domain where the fractal fluctuations dominate, the 3-space coordinates $X^{k}(t, \epsilon_{k}, \epsilon_{t})$ are fractal functions of the time $t$ and the coordinate and time resolutions $\epsilon_k (k = 1,2,3)$ and $\epsilon_t$, respectively. This implies that, for a given $X^{k}$, a positive elementary displacement $dt$ of the curvilinear parameter $t$ induces a displacement $dX^k$ of $X^{k}$ and a negative elementary displacement $-dt$ yields a displacement $-dX^k$, the amplitudes of which are not necessarily equal.

We therefore apply to these two elementary displacements the canonical decomposition of equations (\ref{eq.8})-(\ref{eq.10})
\begin{equation}
dX^k = dx^k + d\xi^k,
\label{eq.11}
\end{equation}
\begin{equation}
dx^k = v^k_{\stackrel{+}{k}} \; dt,
\label{eq.12}
\end{equation}
\begin{equation}
d\xi^k=a^k_+ \sqrt{2 \cal{D}} \, dt^{1/2},
\label{eq.13}
\end{equation}
with $\langle a^k_+ \rangle =0, \langle (a^k_+)^2 \rangle =1$, and
\begin{equation}
-dX^k = {\tilde d}x^k + {\tilde d}\xi^k,
\label{eq.14}
\end{equation}
\begin{equation}
{\tilde d}x^k = v^k_{\stackrel{-}{k}} \; dt,
\label{eq.15}
\end{equation}
\begin{equation}
{\tilde d}\xi^k=a^k_- \sqrt{2 \cal{D}} \, dt^{1/2},
\label{eq.16}
\end{equation}
with $\langle a^k_- \rangle =0, \langle (a^k_-)^2 \rangle =1$.

In the differentiable case, $dX^k=-(-dX^k)$, and therefore 
$v^{k}_{\stackrel{+}{k}}=-v^{k}_{\stackrel{-}{k}}$. This is no longer the case in the non-differentiable case, where the local symmetry $dX^{k}\leftrightarrow -dX^{k}$ is broken. 

This new symmetry breaking should now be combined with the previously studied one, namely the breaking of the symmetry $dt \leftrightarrow -dt$, proceeding from the twofold definition of the derivative with respect to the curvilinear parameter $t$. An elementary displacement $dt$ gives two `classical' derivatives $d/dt_+$ and $d/dt_-$, which, applied to $X^k$, yield in turn two `classical' velocities, $v^k_{{ \stackrel{\pm}{t}} {\stackrel{+}{k}}}$. The same process, applied to an elementary displacement $-dt$, leads us to define again two classical velocities, denoted  by $v^k_{{\stackrel{\pm}{t}} {\stackrel{-}{k}}}$. We summarize this result as 
\begin{equation}
v^{k}_{{\stackrel{\pm}{t}} {\stackrel{+}{k}}}={\frac{dx^k} {dt_{\pm}}}, \qquad 
v^{k}_{{\stackrel{\pm}{t}} {\stackrel{-}{k}}}={\frac{{\tilde d}x^k} {dt_{\pm}}} \; .
\label{eq.17}
\end{equation}

Contrary to what happens in the differentiable case, the total derivative with respect to time of a fractal function $f\left[X^k(t,dt),t,dt\right]$ of integer fractal dimension contains finite terms up to the higher order \cite{AE05}. For a fractal dimension $D_F=2$, the total derivative is written as
\begin{equation}
{\frac{df} {dt}}= {\frac{\partial f} {\partial t}} + {\frac{\partial f}{\partial X^k}}{\frac{dX^k}{dt}} + {\frac{1}{2}}{\frac{\partial^2 f}{\partial X^j \partial X^k}}{\frac{dX^jdX^k}{dt}}.
\label{eq.18}
\end{equation}
We can, at this stage, define several different total derivatives of this fractal function $f$ with respect to time $t$. We write them, using a compact straightforward notation with summation over repeated indices,
\begin{equation}
{\frac{df}{dt}}_{{\stackrel{\pm}{t}} {\stackrel{\pm}{x}} { \stackrel{\pm}{y}} {\stackrel{\pm}{z}}} = {\frac{\partial f} {\partial t}} + (v^{k}_{{\stackrel{\pm}{t}} {\stackrel{\pm}{k}}} + w^{k}_{{\stackrel{\pm}{t}} {\stackrel{\pm}{k}}}){\frac{\partial f} {\partial X^k}} + a^j_\pm a^k_\pm 
{\cal D}{\frac{\partial^2 f} {dX^j dX^k}} \; ,
\label{eq.19}
\end{equation}
with
\begin{equation}
w^k=a^k {\sqrt {2 {\cal D}}} \, dt^{-{\frac{1}{2}}}.
\label{eq.20}
\end{equation}

Now, when we take the stochastic mean of the total derivative of $f$, since  $\langle w^k \rangle =0$ and since
\begin{equation}
\langle d\xi^j_\pm d\xi^k_\pm \rangle = \pm 2 {\cal D} \, \delta^{jk}  \, dt,
\label{eq.21}
\end{equation}
we obtain
\begin{equation}
\langle{\frac{df}{dt}}\rangle_{{\stackrel{\pm}{t}} {\stackrel{\pm}{x}} {\stackrel{\pm}{y}} {\stackrel{\pm}{z}}} = \left({\frac{\partial} {\partial t}} + v^{k}_{{\stackrel{\pm}{t}} {\stackrel{\pm}{k}}} {\frac{\partial}{\partial X^k}} \pm {\cal D}{\Delta}\right) f \; ,
\label{eq.22}
\end{equation}
where the $\pm$ sign in the right-hand side is the same as the $t$-sign. 

By applying these derivatives to the position vector $X^k$, we obtain, as expected,
\begin{equation}
{\frac{dX^{k}}{dt}}_{{\stackrel{\pm}{t}} {\stackrel{\pm}{k}}} = v^{k}_{{\stackrel{\pm}{t}} {\stackrel{\pm}{k}}} .
\label{eq.23}
\end{equation}


\subsection{Covariant derivative operator}


In the simplest case, the breaking of the symmetry $dX^{k}\leftrightarrow -dX^{k}$ is isotropic with respect to the 3-space coordinates (i.e., the signs corresponding to the three $k$ indices are chosen equal). We are left with four non-degenerate 
components $v^{k}_{{\stackrel{\pm}{t}} {\stackrel{\pm}{k}}}$, which we use to define a quaternionic velocity. In its symplectic form, it reads
\begin{equation}
{\cal V}^k={\frac{1}{2}}(v^k_{++} + v^k_{--})-{\frac{i}{2}}
(v^k_{++} - v^k_{--}) +\left[{\frac{1}{2}}(v^k_{+-} + 
v^k_{-+})+{\frac{i}{2}}(v^k_{+-} - v^k_{-+})\right] j.
 \label{eq.24}
\end{equation}
The zero-spin case can be easily recovered as a particular case of this expression. Indeed, it corresponds to a symmetry breaking $dt \leftrightarrow -dt$ (which yields, as recalled hereafter, a complex wavefunction) and no $dx \leftrightarrow -dx$ symmetry breaking. We are left in this case with one doubling only, so that the $j$-term in equation~(\ref{eq.24}) disappears and $v_{++} =v_+, v_{--} =v_-$. This yields in turn the complex ${\cal V}^k$ expression used for the derivation of the Schr\"odinger equation \cite{LN93},
\begin{equation}
{\cal V}^k={\frac{1}{2}}(v^k_{+} + v^k_{-})-{\frac{i}{2}}
(v^k_{+} - v^k_{-}).
\label{eq.25}
\end{equation}
The quaternionic velocity defined such as in equation~(\ref{eq.24}) corresponds to a quaternionic derivative operator $\dfr /dt$ similarly defined, 
\begin{align}
 {\frac{\dfr}{dt}}={\frac{1}{2}}\left({\frac{d}{dt}}_{++}+{\frac{d} 
{dt}}_{--}\right) - {\frac{i}{2}}\left({\frac{d}{dt}}_{++} - {\frac{d}{dt}}_{--}\right) \\ \nonumber
+ \left[{\frac{1}{2}}\left({\frac{d}{dt}}_{+-} + {\frac{d}{dt}}_{-+}\right) + {\frac{i}{2}}\left({\frac{d}{dt}}_{+-} - {\frac{d}{dt}}_{-+}\right)\right] j,
\label{eq.26}
\end{align}
and yielding, when applied to the position vector $X^k$, the corresponding velocity ${\cal V}^k$. We substitute equation~(\ref{eq.22}) into equation~(26) 
and obtain the quaternionic operator
\begin{equation}
{\frac{\dfr}{dt}}= \frac{\partial}{\partial t}+ {\cal V}. \nabla - 
i {\cal D} \Delta \; .
\label{eq.27}
\end{equation}
We recall that the transition from classical (differentiable) mechanics to the scale relativistic one is implemented by replacing the standard time derivative $d/dt$ by the operator $\dfr/dt$ \cite{LN93,CN04} (while accounting for the fact, in particular when using the Leibniz rule, that it is a linear combination of first order and second order derivatives, see e.g. \cite{JCP99,LN04}). This means that $\dfr/dt$ plays the role of a `covariant derivative operator', i.e., of a tool that preserves the form invariance of the equations.


\subsection{Geodesics equation in a fractal space}


Now, we go on with a scale-relativistic-like construction of the corresponding motion equation by generalizing standard classical mechanics using this covariance and a geodesics principle. A quaternionic Lagrange function and the corresponding quaternionic action are obtained from the classical Lagrange function $L (x, v, t)$ and the classical action $S$ by replacing $d/dt$ by $\dfr/dt$. The stationary action principle applied to this quaternionic action yields generalized Euler-Lagrange equations. We also define a generalized quaternionic momentum which satisfies a well-known relation of mechanics \cite{LN96,CN04}
\begin{equation}
{\cal P} = - \nabla {\cal S},
\label{eq.28}
\end{equation}
which can be written as
\begin{equation}
{\cal P} = m {\cal V} ,
\label{eq.29}
\end{equation}
such that, from equation~(\ref{eq.28}), the quaternionic velocity ${\cal V}$ appears as the gradient of the quaternionic action
\begin{equation}
{\cal V} = - \nabla {\cal S}/ m.
\label{eq.30}
\end{equation}
We now introduce a quaternionic wavefunction $\psi$ which is nothing but another expression for the quaternionic action ${\cal S}$, namely
\begin{equation}
\psi^{-1} \nabla \psi = {\frac{i}{S_0}} \nabla {\cal S},
\label{eq.31}
\end{equation}
where $S_0$ is a constant which is introduced for dimensional reasons. It yields, for the quaternionic velocity, as derived from equation~(\ref{eq.30}),
\begin{equation}
{\cal V}=i{\frac{S_0}{m}} \, \psi^{-1} \nabla \psi.
\label{eq.32}
\end{equation} 
Using the generalized covariance principle implemented by the covariant derivative of equation~(\ref{eq.27}), we can now write the equation of motion under the form of a geodesics equation, i.e., of a free-like motion equation,
\begin{equation}
{\frac{\dfr \, {\cal V}}{dt}} = 0.
\label{eq.33}
\end{equation} 
By substituting, in this equation, the expressions of $\dfr/dt$ and ${\cal V}$ as given by equations~(\ref{eq.27}) and (\ref{eq.32}), we obtain
\begin{equation}
{\frac{\partial}{\partial t}} \left(\psi^{-1}.\nabla \psi\right) - i{\frac{S_0}{m}} \psi^{-1}\nabla \psi . \nabla \left(\psi^{-1}.\nabla \psi\right) - i {\cal D} \Delta \left(\psi^{-1}.\nabla \psi\right) = 0
\label{eq.34}
\end{equation} 
The constant  $S_0$ has been proved to be given in a general way (without any assumption), in the Schr\"odinger case, by the relation ${S}_0 = 2 m {\cal D}$ \cite{LN05}. Now, since the Schr\"odinger equation is the limit of the Pauli equation for a particle without spin, this relation must remain valid in the Pauli case. It gives
\begin{equation}
{\frac{\partial}{\partial t}} \left(\psi^{-1}.\nabla \psi\right) - 2i{\cal D} \left[\psi^{-1}\nabla \psi . \nabla \left(\psi^{-1}.\nabla \psi\right) + {\frac{1}{2}} \Delta \left(\psi^{-1}.\nabla \psi\right)\right] = 0
\label{eq.35}
\end{equation}
The definition of the inverse of a quaternion
\begin{equation}
\psi \, \psi^{-1} =  \psi^{-1} \psi = 1,
\label{eq.36}
\end{equation}
implies that $\psi$ and $\psi^{-1}$ commute. But this is not necessarily the case for $\psi$ and $\nabla \psi^{-1}$ nor for $\psi^{-1}$ and $\nabla \psi$. However, when we differentiate equation~(\ref{eq.36}) with respect to the coordinates, we obtain
\begin{equation}
\psi \;.\; \nabla \psi^{-1} = - \nabla \psi \;.\; \psi^{-1} ,
\qquad \qquad
\psi^{-1} \;.\; \nabla \psi = - \nabla \psi^{-1} \;.\; \psi .
\label{eq.37}
\end{equation}

Developing equation~(\ref{eq.35}), using equations~(\ref{eq.37}) and the property $\Delta \nabla = \nabla \Delta$, we obtain, after some calculations,
\begin{equation}
{\frac{\partial}{\partial t}} \left(\psi^{-1}.\nabla \psi\right) -2i{\cal D}\nabla\left[\Delta \psi \;.\; \psi^{-1}\right] = 0.
\label{eq.38}
\end{equation}

Contrary to what happens in the calculations which allowed us to derive the Schr\"odinger \cite{LN93}, Klein-Gordon\cite{LN94} and Dirac \cite{CN04} equations, equation~(\ref{eq.38}) is not a gradient and therefore it is not integrable. It remains a third-order equation, the physical meaning of which is not obvious. It cannot be put under the Pauli form, even with the addition of an electromagnetic field. Since, in the scale relativity theory, the meaning of the function $\psi$ is not set as an axiom (as in standard quantum mechanics) but instead deduced from the developments of the formalism (concerning in particular its status of wavefunction and the proof of Born's postulate \cite{CN04,LN05}), the impossibility to put this equation in the Pauli form also deprives such a function $\psi$ (constructed by taking into account the $dx \leftrightarrow -dx$ symmetry breaking in fractal space instead of full fractal spacetime) of its physical meaning.

This apparent failure is actually a success of the scale relativity theory. Indeed, it is in accordance with the results obtained in the framework of standard quantum mechanics, but it also enlightens it in a new way, by clearly establishing (in this framework) the double non-differentiable and relativistic origin of spin, even when it manifests itself in a non-relativistic (low energy) situation.

Since the attempt to extend the Schr\"odinger equation to spinors failed, Pauli's equation must be derived from Dirac's, in the scale relativity approach to quantum mechanics also. It therefore yields the right value for the electron gyromagnetic factor. We shall discuss this point more thoroughly at the end of section \ref{s.ped}.


\section{The Pauli equation as a non-relativistic limit of the Dirac equation in the quaternionic formalism}
\label{s.ped}


The Dirac equation for a free fermion has been derived, in the framework of scale relativity, as a mere square root of the free Klein-Gordon equation, written whith a bi-quaternionic wavefunction thanks to three successive doublings of its mathematical representation \cite{CN03,CN04}. Its covariant form is the usual
\begin{equation}
\left(i\hbar \gamma^{\mu}\partial_{\mu} - mc\right) \psi = 0.
\label{eq.39}
\end{equation}
A scale relativistic theory of electromagnetism has also been developed \cite{LN96,LN94,LN03} in which the charges are built from the symmetries of the `scale space', according to Noether's theorem, and the electromagnetic field is linked to the resolution dilation of the internal structures of the fractal spacetime \footnote {The formalism applied to the Abelian gauge theory of electromagnetism has subsequently been extended to non-Abelian gauge theories, whose tools have now been given physical meanings in the framework of scale relativity \cite{NCL06}.}. This approach has allowed us to establish, from the first principles of the theory, the form of the action as it appears in standard electromagnetism and, in particular, the form of the particle-field coupling term (which is postulated in the standard theory).

We can therefore write the Dirac equation for an electron of mass $m$ and charge $e$, in an external electromagnetic field $A_{\mu}$, as
\begin{equation}
\left[\gamma^{\mu}\left(i\hbar \partial_{\mu} - e A_{\mu}\right) - mc\right] \psi = 0 \; ,
\label{eq.40}
\end{equation}
with the wavefunction $\psi$ under the form of a bi-quaternion (complex quaternion)
\begin{equation}
\psi = \psi_0 + i\psi_1 + j\psi_2 +  k\psi_3 .
\label{eq.41}
\end{equation}
This mathematical representation of the wavefunction is equivalent to the spinorial one, with the following correspondence of the components
\begin{equation}
\psi = \left (
\begin{array}{c}
\psi_0 \\
\psi_1 \\
\psi_2 \\
\psi_3 \\
\end{array}
\right ). 
\label{eq.42}
\end{equation}
We can now derive the Pauli equation, following the standard method of eliminating small components.

We consider a two-component representation, where the four-component spinor $\psi$ is decomposed into two two-component spinors $\phi$ and $\chi$.
\begin{equation}
\psi = \left (
\begin{array}{c}
\phi \\
\chi \\
\end{array}
\right ). 
\label{eq.43}
\end{equation}
In the quaternionic representation, this amounts to shifting to a symplectic form of $\psi$, where
\begin{equation}
\phi = \psi_0 + i\psi_1 ,
\label{eq.44}
\end{equation}
and
\begin{equation}
\chi = \psi_2 - i\psi_3 ,
\label{eq.45}
\end{equation}
In the non-relativistic limit, the rest energy, $mc^2$, becomes dominant. Therefore, the two-component solution is approximately
\begin{equation}
\left ( \begin{array}{c}
\phi \\
\chi \\
\end{array}
\right ) = 
\left (
\begin{array}{c}
\phi' \\
\chi' \\
\end{array}
\right )
e^{-imc^2t/\hbar}, 
\label{eq.46}
\end{equation}
where $\phi'$ and $\chi'$ are slowly varying functions of time.  Substitution of this non-relativistic solution into the Dirac equation, equation~(\ref{eq.40}), in the Dirac representation, gives
\begin{eqnarray}
i\hbar {\frac{\partial \phi'}{\partial t}} &=& c \overrightarrow{\sigma} . \left(i\hbar \nabla - {\frac{e}{c}} \overrightarrow{A}\right)\chi' + e A_0 \phi' , \\
\label{eq.47}
i\hbar {\frac{\partial \chi'}{\partial t}} &=&  c \overrightarrow{\sigma} . \left(i\hbar \nabla - {\frac{e}{c}} \overrightarrow{A}\right)\phi' + e A_0 \chi' - 2mc^2 \chi' .
\label{eq.48}
\end{eqnarray}
When the kinetic energy is small compared to the rest energy, $\chi'$ is a slowly varying function of time and
\begin{equation}
\left|i\hbar {\frac{\partial \chi'}{\partial t}}\right| \ll |mc^2 \chi'| .
\label{eq.49}
\end{equation}
When the electrostatic potential $A_0$ is weak, the potential energy is small compared to the rest energy.
\begin{equation}
|e A_0 \chi'| \ll |mc^2 \chi'| .
\label{eq.50}
\end{equation}
With these last two approximations, equation~(\ref{eq.48}) becomes
\begin{equation}
0 \approx c \overrightarrow{\sigma} . \left(i\hbar \nabla - {\frac{e}{c}} \overrightarrow{A}\right)\phi' - 2 mc^2 \chi' ,
\label{eq.51}
\end{equation}
which gives
\begin{equation}
\chi' = {\frac{\overrightarrow{\sigma} . \left(i\hbar \nabla - {\frac{e}{c}} \overrightarrow{A}\right)}{2mc}} \phi' .
\label{eq.52}
\end{equation}
The lower component, $\chi$, is generally refered to as the `small' component of the wavefunction $\psi$, relative to the `large' component, $\phi$. The small component is approximately $v/c$ less than the large one in the non-relativistic limit.

Substituting the expression for $\chi'$, given by equation~(\ref{eq.52}), into equation (47), we obtain
\begin{equation}
i\hbar {\frac{\partial \phi'}{\partial t}} = {\frac {\overrightarrow{\sigma} . \left(i\hbar \nabla - {\frac{e}{c}} \overrightarrow{A}\right) \overrightarrow{\sigma} . \left(i\hbar \nabla - {\frac{e}{c}} \overrightarrow{A}\right)}{2m}} \phi' + e A_0 \phi'.
\label{eq.53}
\end{equation}
We are therefore left with an equation for the 2-spinor $\phi'$ alone which, in the quaternionic formalism, is written as
\begin{equation}
\phi' = (\psi_0 + i\psi_1)e^{imc^2 t /\hbar} .
\label{eq.54}
\end{equation}
Finally, by using the well-known identities,
\begin{equation}
(\overrightarrow{\sigma} \overrightarrow{a}) (\overrightarrow{\sigma} \overrightarrow{b}) = \overrightarrow{a} \overrightarrow{b} + i \overrightarrow{\sigma}(\overrightarrow{a} \times \overrightarrow{b}) ,
\label{eq.55}
\end{equation}
\begin{equation}
(\nabla \times \overrightarrow{A} + \overrightarrow{A} \times \nabla ) \phi' = \mathrm{curl} \overrightarrow{A},
\label{eq.56}
\end{equation}
we obtain, $\overrightarrow{B} = \mathrm{curl} \overrightarrow{A}$ being the magnetic field,
\begin{equation}
i\hbar {\frac{\partial \phi'}{\partial t}} = \left[{\frac{1}{2m}} \left(i\hbar \nabla - {\frac{e}{c}} \overrightarrow{A}\right)^2 - {\frac{e \hbar}{2mc}} \overrightarrow{\sigma} . \overrightarrow{B} + e A_0 \right] \phi'.
\label{eq.57}
\end{equation}
We recognize here the Pauli equation for the theory of spin in non-relativistic quantum mechanics, with $i\hbar \nabla$ replacing the momentum operator $\hat{p}$, implementing the correspondence principle. Note that this correspondence principle for the momentum can be derived in the framework of scale relativity \cite{LN93} and needs no more to be postulated as in standard quantum mechanics. As it is well known, one of the main results of the Pauli equation (when it is derived from the Dirac equation) is to yield the correct gyromagnetic factor $g = 2$ for a free electron.

We have therefore shown that the scale relativity approach agrees with the physics of the Pauli bi-spinors for spin-1/2 particles. Since the correct motion equation for these spinors cannot be obtained from the mere symmetry breakings $dt \leftrightarrow -dt$ and $dx \leftrightarrow -dx$ in a fractal 3-space, this confirms and reinforces the relativistic nature of spin.

The difference between the two approaches of sections \ref{s.psb} and \ref{s.ped} is that, in the first one, we consider the time $t$ as a curvilinear parameter on the geodesics of a fractal 3-space and, in the second one, we consider it as a coordinate of a fractal 4-spacetime. Since we are looking for motion equations, we cannot get rid of the time dependence and we are left with a non-integrable time-dependent term in equation~(\ref{eq.38}). In contrast, when dealing with the Dirac equation, we choose, as a curvilinear parameter, the relativistic line element $s$, which we need not retain in the equation for a first approach. We thus obtain an integrable equation of which the quaternionic square root is the Dirac equation \cite{CN03,CN04}. But here, time is considered as one of the four coordinates of the underlying fractal spacetime and remains such in the equations, namely in the Dirac equation, and therefore also in the Pauli equation.


\section{The Pauli quaternionic velocity as a degeneracy of the Dirac bi-quaternionic one}
\label{s.srp}


For the derivation of the Dirac equation in the framework of scale relativity that was developed in \cite{CN03,CN04}, we have chosen a peculiar expression for ${\cal V}^{\mu}$ as a function of the eight non-degenerate components of the classical velocity, $v^{\mu}_{\pm \pm}$ and ${\tilde v}^{\mu}_{\pm \pm}$,
 \begin{eqnarray}
{\cal V}^\mu={\frac{1}{2}}(v^\mu_{++} + {\tilde v}^\mu_{--})-{\frac{i}{2}}(v^\mu_{++} - {\tilde v}^\mu_{--}) +\left[{\frac{1}{2}}(v^\mu_{+-} + 
v^\mu_{-+})-{\frac{i}{2}}(v^\mu_{+-} - {\tilde v}^\mu_{++})\right] e_1 
\nonumber \\
+ \left[{\frac{1}{2}}(v^\mu_{--} + {\tilde v}^\mu_{+-})-{\frac{i}{2}}(v^\mu_{--} - 
{\tilde v}^\mu_{-+})\right] e_2 + \left[{\frac{1}{2}}(v^\mu_{-+} + 
{\tilde v}^\mu_{++})-{\frac{i}{2}}({\tilde v}^\mu_{-+} + 
{\tilde v}^\mu_{+-})\right] e_3 .
\label{eq.58}
\end{eqnarray}
Here $i$ denotes the imaginary element used to write the complex components of the bi-quaternions as sums of a real and an imaginary part and $e_i$ s, with $i= 1,2,3$, correspond to the $i,j,k$ imaginary elements used above to write linear expressions of the quaternions. We can therefore write any bi-quaternionic wavefunction, solution of the Dirac equation, as a function of its eight real components,
\begin{equation}
\psi = \phi_0 + i \chi_0 + (\phi_1 + i \chi_1)e_1 + (\phi_2 + i \chi_2)e_2 + (\phi_3 + i \chi_3)e_3.
\label{eq.59}
\end{equation}
Since $\psi$ is assimilated to a Dirac bi-spinor, it must have the same properties. In particular, it must be a unit quaternion, i.e., a quaternion with a unit norm, which is written as
\begin{equation}
\phi_0^2 + \chi_0^2 + \phi_1^2 + \chi_1^2 + \phi_2^2 + \chi_2^2 + \phi_3^2 + \chi_3^2 = 1.
\label{eq.59b}
\end{equation}
In the motion-relativistic case of the Dirac equation, the bi-quaternionic action is given by
\begin{equation}
d{\cal S} = \partial _\mu {\cal S} \, d x^\mu = -mc{\cal V}_\mu dx^\mu.
\label{eq.59c}
\end{equation}
The bi-quaternionic 4-momentum is therefore
\begin{equation}
{\cal P}_\mu = mc{\cal V}_\mu = -\partial _\mu {\cal S}.
\label{eq.59d}
\end{equation}
We can then introduce a bi-quaternionic wavefunction, defined as
\begin{equation}
\psi^{-1} \partial_\mu \psi = {\frac{i}{{\cal S}_0}} \partial _\mu {\cal S}.
\label{eq.59e}
\end{equation}
Now, from equation~(\ref{eq.59}), we calculate the components of the bi-quaternionic 4-velocity,
\begin{equation}
{\cal V}_{\mu}=i{\frac{S_0}{mc}} \psi^{-1} \partial_{\mu} \psi.
\label{eq.60}
\end{equation}
(We correct here a misprint of the equivalent expression in \cite{CN04} where $c$ was lacking). We obtain:
\begin{multline}
{\cal V}^\mu={\frac{{\cal S}_0} {mc}} [-(\phi_0\partial^\mu\chi_0 + \chi_0\partial^\mu\phi_0 + \phi_1\partial^\mu\chi_1 + \chi_1\partial^\mu\phi_1 +  \phi_2\partial^\mu\chi_2 + \chi_2\partial^\mu\phi_2 + \phi_3\partial^\mu\chi_3 + \chi_3\partial^\mu\phi_3) \\
+ ( \phi_0\partial^\mu\phi_0 - \chi_0\partial^\mu\chi_0 + \phi_1\partial^\mu\phi_1 - \chi_1\partial^\mu\chi_1 +  \phi_2\partial^\mu\phi_2 - \chi_2\partial^\mu\chi_2 + \phi_3\partial^\mu\phi_3 - \chi_3\partial^\mu\chi_3)i \\ 
+[-\phi_0\partial^\mu\chi_1 - \chi_0\partial^\mu\phi_1 + \phi_1\partial^\mu\chi_0 + \chi_1\partial^\mu\phi_0 +  \phi_2\partial^\mu\chi_3 + \chi_2\partial^\mu\phi_3 - \phi_3\partial^\mu\chi_2 - \chi_3\partial^\mu\phi_2 \\ 
+ (\phi_0\partial^\mu\phi_1 - \chi_0\partial^\mu\chi_1 - \phi_1\partial^\mu\phi_0 + \chi_1\partial^\mu\chi_0 -  \phi_2\partial^\mu\phi_3 + \chi_2\partial^\mu\chi_3 + \phi_3\partial^\mu\phi_2 - \chi_3\partial^\mu\chi_2)i  ] e_1 \\ 
+[-\phi_0\partial^\mu\chi_2 - \chi_0\partial^\mu\phi_2 - \phi_1\partial^\mu\chi_3 - \chi_1\partial^\mu\phi_3 +  \phi_2\partial^\mu\chi_0 + \chi_2\partial^\mu\phi_0 + \phi_3\partial^\mu\chi_1 + \chi_3\partial^\mu\phi_1 \\ 
+ ( \phi_0\partial^\mu\phi_2 - \chi_0\partial^\mu\chi_2 + \phi_1\partial^\mu\phi_3 - \chi_1\partial^\mu\chi_3 -  \phi_2\partial^\mu\phi_0 + \chi_2\partial^\mu\chi_0 - \phi_3\partial^\mu\phi_1 + \chi_3\partial^\mu\chi_1)i  ]e_2 \\
+[-\phi_0\partial^\mu\chi_3 - \chi_0\partial^\mu\phi_3 + \phi_1\partial^\mu\chi_2 + \chi_1\partial^\mu\phi_2 -  \phi_2\partial^\mu\chi_1 - \chi_2\partial^\mu\phi_1 + \phi_3\partial^\mu\chi_0 + \chi_3\partial^\mu\phi_0 \\
+ ( \phi_0\partial^\mu\phi_3 - \chi_0\partial^\mu\chi_3 - \phi_1\partial^\mu\phi_2 + \chi_1\partial^\mu\chi_2 +  \phi_2\partial^\mu\phi_1 - \chi_2\partial^\mu\chi_1 - \phi_3\partial^\mu\phi_0 + \chi_3\partial^\mu\chi_0)i ]e_3].
\label{eq.61}
\end{multline}
Then, we identify the real term and the imaginary terms of the same kind in equations~(\ref{eq.58}) and (\ref{eq.61}) and obtain eight equations giving linear combinations of the $v^{\mu}_{\pm \pm}$ and ${\tilde v}^{\mu}_{\pm \pm}$ as functions of the components $\phi_i$ and $\chi_i$ of the wavefunction and their derivatives. Taking linear combinations of the first two of these equations, involving $v^{\mu}_{++}$ and ${\tilde v}^{\mu}_{--}$, we obtain 
\begin{multline}
v^{\mu}_{++} = - {\frac{{\cal S}_0}{mc}} (
\phi_0\partial^\mu\chi_0 + \chi_0\partial^\mu\phi_0 + \phi_1\partial^\mu\chi_1 + \chi_1\partial^\mu\phi_1 +  \phi_2\partial^\mu\chi_2 + \chi_2\partial^\mu\phi_2 + \phi_3\partial^\mu\chi_3 + \chi_3\partial^\mu\phi_3
 \\
+ \phi_0\partial^\mu\phi_0 - \chi_0\partial^\mu\chi_0 + \phi_1\partial^\mu\phi_1 - \chi_1\partial^\mu\chi_1 +  \phi_2\partial^\mu\phi_2 - \chi_2\partial^\mu\chi_2 + \phi_3\partial^\mu\phi_3 - \chi_3\partial^\mu\chi_3)
\label{eq.62}
\end{multline}
and
\begin{multline}
{\tilde v}^{\mu}_{--} = - {\frac{{\cal S}_0}{mc}} (
\phi_0\partial^\mu\chi_0 + \chi_0\partial^\mu\phi_0 + \phi_1\partial^\mu\chi_1 + \chi_1\partial^\mu\phi_1 +  \phi_2\partial^\mu\chi_2 + \chi_2\partial^\mu\phi_2 + \phi_3\partial^\mu\chi_3 + \chi_3\partial^\mu\phi_3
 \\
- \phi_0\partial^\mu\phi_0 + \chi_0\partial^\mu\chi_0 - \phi_1\partial^\mu\phi_1 + \chi_1\partial^\mu\chi_1 -  \phi_2\partial^\mu\phi_2 + \chi_2\partial^\mu\chi_2 - \phi_3\partial^\mu\phi_3 + \chi_3\partial^\mu\chi_3).
\label{eq.63}
\end{multline}
We have seen in equation (\ref{eq.51}) that the Pauli spinor, $\phi'$, is a function of the components $\phi_0$, $\chi_0$, $\phi_1$ and $\chi_1$ only, the other `small' components vanishing at the non-relativistic limit. Therefore, at this limit, we see, from equation~(\ref{eq.63}), that ${\tilde v}^{\mu}_{\pm \pm}$ does not vanish, since it includes terms which are products of `large' components and their derivatives. However, for the contruction of the 4-velocity and the covariant derivative we proposed in \cite{CN03,CN04}, we denoted with a tilde the components issued from the breaking of the time-reversal and parity symmetries. Since the breaking of these symmetries is a mere property of relativistic motion, tilde terms are not supposed to appear in Pauli's non-relativistic velocity. We therefore conclude that the expression we retained in \cite{CN03,CN04} for ${\cal V}^{\mu}$ was not the proper one.

But we could have chosen any other expression for this bi-quaternionic velocity provided it fulfils the requirements that, at the non-relativistic limit, we recover a quaternionic velocity for the  Pauli spinor, a complex one for the Schr\"odinger equation and that, at the classical limit, every term vanishes except the real one. We therefore propose here a more symmetrical expression than that retained in \cite{CN03,CN04},  
\begin{align}
{\cal V}^\mu={\frac{1}{2}}(v^\mu_{++} +  v^\mu_{--})-{\frac{i}{2}}
(v^\mu_{++} - v^\mu_{--}) +\left[{\frac{1}{2}}(v^\mu_{+-} + 
v^\mu_{-+})-{\frac{i}{2}}(v^\mu_{+-} - v^\mu_{-+})\right] e_1 
\nonumber \\
+ \left[{\frac{1}{2}}({\tilde v}^\mu_{++} + {\tilde v}^\mu_{--})-{\frac{i}{2}}({\tilde v}^\mu_{++} - 
{\tilde v}^\mu_{--})\right] e_2 + \left[{\frac{1}{2}}({\tilde v}^\mu_{+-} + {\tilde v}^\mu_{-+}) - {\frac{i}{2}}({\tilde v}^\mu_{+-} - {\tilde v}^\mu_{-+})\right] e_3 .
\label{eq.64}
\end{align}
Note that the development of the theory is independent of this choice, since the covariant derivative keeps the same form whatever the mathematical expression of ${\cal V}^{\mu}$ is.

As previously, we identify the real term and the imaginary terms of the same kind in equations~(\ref{eq.61}) and (\ref{eq.64}) and get eight equations giving linear combinations of the $v^{\mu}_{\pm \pm}$ and ${\tilde v}^{\mu}_{\pm \pm}$ as functions of the components $\phi_i$ and $\chi_i$ of the Dirac wavefunction and their derivatives. Taking linear combinations of these equations, we obtain
\begin{multline}
v^{\mu}_{++} = - {\frac{{\cal S}_0}{mc}} (
\phi_0\partial^\mu\chi_0 + \chi_0\partial^\mu\phi_0 + \phi_1\partial^\mu\chi_1 + \chi_1\partial^\mu\phi_1 +  \phi_2\partial^\mu\chi_2 + \chi_2\partial^\mu\phi_2 + \phi_3\partial^\mu\chi_3 + \chi_3\partial^\mu\phi_3
 \\
+ \phi_0\partial^\mu\phi_0 - \chi_0\partial^\mu\chi_0 + \phi_1\partial^\mu\phi_1 - \chi_1\partial^\mu\chi_1 +  \phi_2\partial^\mu\phi_2 - \chi_2\partial^\mu\chi_2 + \phi_3\partial^\mu\phi_3 - \chi_3\partial^\mu\chi_3),
\label{eq.65}
\end{multline}
\begin{multline}
v^{\mu}_{--} = - {\frac{{\cal S}_0}{mc}} (
\phi_0\partial^\mu\chi_0 + \chi_0\partial^\mu\phi_0 + \phi_1\partial^\mu\chi_1 + \chi_1\partial^\mu\phi_1 +  \phi_2\partial^\mu\chi_2 + \chi_2\partial^\mu\phi_2 + \phi_3\partial^\mu\chi_3 + \chi_3\partial^\mu\phi_3
 \\
- \phi_0\partial^\mu\phi_0 + \chi_0\partial^\mu\chi_0 - \phi_1\partial^\mu\phi_1 + \chi_1\partial^\mu\chi_1 -  \phi_2\partial^\mu\phi_2 + \chi_2\partial^\mu\chi_2 - \phi_3\partial^\mu\phi_3 + \chi_3\partial^\mu\chi_3),
\label{eq.66}
\end{multline}
\begin{multline}
v^{\mu}_{+-} = - {\frac{{\cal S}_0}{mc}} (
\phi_0\partial^\mu\chi_1 + \chi_0\partial^\mu\phi_1 - \phi_1\partial^\mu\chi_0 - \chi_1\partial^\mu\phi_0 -  \phi_2\partial^\mu\chi_3 - \chi_2\partial^\mu\phi_3 + \phi_3\partial^\mu\chi_2 + \chi_3\partial^\mu\phi_2
 \\
+ \phi_0\partial^\mu\phi_1 - \chi_0\partial^\mu\chi_1 - \phi_1\partial^\mu\phi_0 + \chi_1\partial^\mu\chi_0 -  \phi_2\partial^\mu\phi_3 + \chi_2\partial^\mu\chi_3 + \phi_3\partial^\mu\phi_2 - \chi_3\partial^\mu\chi_2),
\label{eq.67}
\end{multline}
\begin{multline}
v^{\mu}_{-+} = - {\frac{{\cal S}_0}{mc}} (
\phi_0\partial^\mu\chi_1 + \chi_0\partial^\mu\phi_1 - \phi_1\partial^\mu\chi_0 - \chi_1\partial^\mu\phi_0 -  \phi_2\partial^\mu\chi_3 - \chi_2\partial^\mu\phi_3 + \phi_3\partial^\mu\chi_2 + \chi_3\partial^\mu\phi_2
 \\
- \phi_0\partial^\mu\phi_1 + \chi_0\partial^\mu\chi_1 + \phi_1\partial^\mu\phi_0 - \chi_1\partial^\mu\chi_0 +  \phi_2\partial^\mu\phi_3 - \chi_2\partial^\mu\chi_3 - \phi_3\partial^\mu\phi_2 + \chi_3\partial^\mu\chi_2),
\label{eq.68}
\end{multline} 
\begin{multline}
{\tilde v}^{\mu}_{++} = - {\frac{{\cal S}_0}{mc}} (
\phi_0\partial^\mu\chi_2 + \chi_0\partial^\mu\phi_2 + \phi_1\partial^\mu\chi_3 + \chi_1\partial^\mu\phi_3 -  \phi_2\partial^\mu\chi_0 - \chi_2\partial^\mu\phi_0 - \phi_3\partial^\mu\chi_1 - \chi_3\partial^\mu\phi_1
 \\
+ \phi_0\partial^\mu\phi_2 - \chi_0\partial^\mu\chi_2 + \phi_1\partial^\mu\phi_3 - \chi_1\partial^\mu\chi_3 -  \phi_2\partial^\mu\phi_0 + \chi_2\partial^\mu\chi_0 - \phi_3\partial^\mu\phi_1 + \chi_3\partial^\mu\chi_1),
\label{eq.69}
\end{multline}
\begin{multline}
{\tilde v}^{\mu}_{--} = - {\frac{{\cal S}_0}{mc}} (
\phi_0\partial^\mu\chi_2 + \chi_0\partial^\mu\phi_2 + \phi_1\partial^\mu\chi_3 + \chi_1\partial^\mu\phi_3 -  \phi_2\partial^\mu\chi_0 - \chi_2\partial^\mu\phi_0 - \phi_3\partial^\mu\chi_1 - \chi_3\partial^\mu\phi_1
 \\
- \phi_0\partial^\mu\phi_2 + \chi_0\partial^\mu\chi_2 - \phi_1\partial^\mu\phi_3 + \chi_1\partial^\mu\chi_3 +  \phi_2\partial^\mu\phi_0 - \chi_2\partial^\mu\chi_0 + \phi_3\partial^\mu\phi_1 - \chi_3\partial^\mu\chi_1),
\label{eq.70}
\end{multline}
\begin{multline}
{\tilde v}^{\mu}_{+-} = - {\frac{{\cal S}_0}{mc}} (
\phi_0\partial^\mu\chi_3 + \chi_0\partial^\mu\phi_3 - \phi_1\partial^\mu\chi_2 - \chi_1\partial^\mu\phi_2 +  \phi_2\partial^\mu\chi_1 + \chi_2\partial^\mu\phi_1 - \phi_3\partial^\mu\chi_0 - \chi_3\partial^\mu\phi_0
\\
+ \phi_0\partial^\mu\phi_3 - \chi_0\partial^\mu\chi_3 - \phi_1\partial^\mu\phi_2 + \chi_1\partial^\mu\chi_2 +  \phi_2\partial^\mu\phi_1 - \chi_2\partial^\mu\chi_1 - \phi_3\partial^\mu\phi_0 + \chi_3\partial^\mu\chi_0)
\label{eq.71}
\end{multline}
and
\begin{multline}
{\tilde v}^{\mu}_{-+} = - {\frac{{\cal S}_0}{mc}} (
\phi_0\partial^\mu\chi_3 + \chi_0\partial^\mu\phi_3 - \phi_1\partial^\mu\chi_2 - \chi_1\partial^\mu\phi_2 +  \phi_2\partial^\mu\chi_1 + \chi_2\partial^\mu\phi_1 - \phi_3\partial^\mu\chi_0 - \chi_3\partial^\mu\phi_0
 \\
- \phi_0\partial^\mu\phi_3 + \chi_0\partial^\mu\chi_3 + \phi_1\partial^\mu\phi_2 - \chi_1\partial^\mu\chi_2 -  \phi_2\partial^\mu\phi_1 + \chi_2\partial^\mu\chi_1 + \phi_3\partial^\mu\phi_0 - \chi_3\partial^\mu\chi_0).
\label{eq.72}
\end{multline}
We can now get rid of all the `small' components and their derivatives, which are all the terms with an index 2 or 3, since these components do not appear in the final Pauli equation (indeed, this equation is the Dirac equation written with the `large' terms alone). We see, from equations (\ref{eq.65})-(\ref{eq.72}), that all the tilde components of the velocity vanish and that the only non-zero ones are the four $v_{++}$, $v_{+-}$, $v_{-+}$ and $v_{--}$. 

We have stated, in section \ref{s.ped}, that the Pauli spinor, $\phi'$, is obtained from the Dirac bi-spinor, $\psi$, as written in equation (\ref{eq.59}), by the correspondence
\begin{equation}
\phi' = (\psi_0 + e_1 \psi_1) \, e^{imc^2t/\hbar} = \left[ \phi_0 + i \chi_0 + (\phi_1 + i \chi_1)e_1 \right]e^{imc^2t/\hbar}.
\label{eq.74}
\end{equation}
This implies that we obtain the quaternionic velocity corresponding to the non-relativistic limit, first, by neglecting the small components in the expression of the bi-quaternionic velocity of equation~(\ref{eq.61}), which gives
\begin{multline}
{\cal V}^\mu={\frac{{\cal S}_0} {mc}}[-(\phi_0\partial^\mu\chi_0 + \chi_0\partial^\mu\phi_0 + \phi_1\partial^\mu\chi_1 + \chi_1\partial^\mu\phi_1)
+ ( \phi_0\partial^\mu\phi_0 - \chi_0\partial^\mu\chi_0 + \phi_1\partial^\mu\phi_1 - \chi_1\partial^\mu\chi_1)i \\
+[-\phi_0\partial^\mu\chi_1 - \chi_0\partial^\mu\phi_1 + \phi_1\partial^\mu\chi_0 + \chi_1\partial^\mu\phi_0
+ (\phi_0\partial^\mu\phi_1 - \chi_0\partial^\mu\chi_1 - \phi_1\partial^\mu\phi_0 + \chi_1\partial^\mu\chi_0)i ]e_1 ],
\label{eq.75}
\end{multline}
then, by replacing in this expression $\phi$ s and $\chi$ s by their primed counterparts as given by equation~(\ref{eq.74}) and the simplectic decomposition
\begin{equation}
\phi' = \left[\phi'_0 + i \chi'_0 + (\phi'_1 + i \chi'_1)e_1 \right].
\label{eq.76}
\end{equation}
For that, we have to distinguish the component ${\cal V}^0$ of the 4-velocity (which, at the non-relativistic limit, must be the light velocity $c$) from its three spatial components which exhibit a different behaviour.

From equations~(\ref{eq.74}) and (\ref{eq.76}), we can write
\begin{equation}
\phi_{0,1} , \chi_{0,1} = \phi'_{0,1} , \chi'_{0,1} e^{-imc^2t/\hbar}.
\label{eq.77}
\end{equation}
We therefore obtain, for the partial derivatives with respect to time of the components of $\psi$,
\begin{equation}
\partial^0 \phi_{0,1} = \partial^0 \phi'_{0,1} e^{-imc^2t/\hbar} - {\frac{imc^2}{\hbar}}\phi'_{0,1} e^{-imc^2t/\hbar},
\label{eq.78}
\end{equation}
and analogous expressions for $\partial^0 \chi_{0,1}$.

We have seen, in section \ref{s.ped}, that $\phi'$ (and therefore its components) is a slowly varying function of time. The partial derivatives with respect to time on the right hand side of equation~(\ref{eq.78}) thus approximately vanish and this equation becomes
\begin{equation}
\partial^0 \phi_{0,1} = - {\frac{imc^2}{\hbar}}\phi'_{0,1} e^{-imc^2t/\hbar},
\label{eq.79}
\end{equation}
and analogous equations for $\partial^0 \chi_{0,1}$.

Substituting the expressions for $\phi_{0,1} , \chi_{0,1}, \partial^0 \phi_{0,1}$ and $\partial^0 \chi_{0,1}$ into equation~(\ref{eq.75}), we obtain, after some calculations and using the property that, in the standard quantum domain, ${\cal S}_0 = \hbar$,
\begin{equation}
{\cal V}^0 = c e^{-2imc^2t/\hbar}\left[\left(\phi'_0 + i\chi'_0 \right)^2 + \left(\phi'_1 + i\chi'_1 \right)^2 \right],
\label{eq.80}
\end{equation}
which we can write, owing to the fact that the norm of the spinor $\phi$ must be normalized to the unity,
\begin{equation}
{\cal V}^0 = c \left[\left(\psi'_0 e^{-imc^2t/\hbar}\right)^2 + \left(\psi'_1 e^{-imc^2t/\hbar}\right)^2 \right] = c \left(\psi_0^2 + \psi_1^2 \right) = c\left|\phi\right|^2 = c
\label{eq.81}
\end{equation}

Now, we consider the three spatial components, ${\cal V}^k$, k = 1,2,3, of the Dirac velocity reduced to its `large' terms as in the non-relativistic limit. In this case,
\begin{equation}
\partial^k \phi_{0,1} = \partial^k \phi'_{0,1} e^{-imc^2t/\hbar}
\label{eq.82}
\end{equation}
and analogous equations for $\chi$ s.

Suppressing the `small' terms (those with the indices 2 and 3) into the expressions of $v^k_{\pm\pm}$ s given by equations (\ref{eq.65})-(\ref{eq.68}), substituting into them the expressions for $\phi_{0,1} , \chi_{0,1}, \partial^k \phi_{0,1}$ and $\partial^k \chi_{0,1}$ given by equations (\ref{eq.77}) and (\ref{eq.82}) and finally combining the results obtained for $v^k_{\pm\pm}$ s such as to reproduce the expression of ${\cal V}^k$, we get
\begin{equation}
{\cal V}^k = {\frac{1}{2}}(v^k_{++} +  v^k_{--})-{\frac{i}{2}}
(v^k_{++} - v^k_{--}) +\left[{\frac{1}{2}}(v^k_{+-} + 
v^k_{-+})-{\frac{i}{2}}(v^k_{+-} - v^k_{-+})\right] e_1.
\label{eq.83}
\end{equation}
We have therefore obtained the velocity corresponding to the non-relativistic case from the mere degeneracy of the bi-quaternionic relativistic one, in the simplectic form of a real quaternion. This degeneracy yields the non-relativistic velocity as a 3-vector naturally derived from the 4-velocity of the relativistic case. The quaternionic velocity of equation (\ref{eq.83}) has therefore all the properties needed to implement the scale relativistic procedure described in section \ref{s.psb}. The failure of this procedure (namely, in the framework of a fractal space, that leads to non-relativistic quantum mechanics) to give the correct Pauli equation (while a full fractal spacetime description, that leads to relativistic quantum mechanics,  is successful) reinforces the fundamentally relativistic nature of spin as it appears in standard quantum mechanics. This corroborates the relevance of scale relativity for the building from first principles of the quantum postulates and of the quantum mechanical tools and results.


\section{Spin as internal angular momentum of geodesics in a fractal space}
\label{s.gfs}


\subsection{Geometric models of the spin}

The geometric description of quantum physics brought by the scale-relativity / fractal-spacetime approach allows one to give a physical picture of what the spin is. Recall that the spin has been considered, since its discovery, as a physical quantity of pure quantum origin having no classical counterpart. Indeed, assuming an extension of the electron of the order of its classical radius $r_e=\alpha^{-1} \lambda_c$, where $\alpha =1/137.036...$ is the fine structure constant and  $\lambda_c=\hbar/mc$ is its Compton length, an angular momentum $\hbar/2$ would involve a velocity of rotation of its surface of order $\alpha^{-1} c$, which is clearly excluded by special relativity.

The scale relativity theory allows one to suggest a new solution to this fundamental problem. This solution remains non-classical (owing to the fact that an everywhere non-differentiable spacetime is non-classical, as proved by the quantum-mechanical-type behaviour of its geodesics), but it is however a geometric solution.

As recalled in this paper, in the scale relativistic framework, both the complex nature of the wavefunction and the existence of spin have a common origin, namely the fundamental two-valuedness of the derivative (in its generalized definition) coming from non-differentiability. These two successive doublings are naturally accounted for in terms of algebra doublings (see the appendix of ~\cite{CN04}), i.e., of a description tool that jumps from real numbers ${\rm I\! R}$ to complex numbers ${\rm I\! \!\!C}={\rm I\! R}^2$, then to quaternions ${\rm I\! H}={\rm I\! \!\!C}^2$. However, while the origin of the complex nature of the wavefunction is linked to the total derivative (and therefore to proper time) through the doubling ${d}/{ds}\rightarrow ({d_+}/{ds},{d_-}/{ds})$, the origin of spin is linked to the partial derivative with respect to the coordinates through the doubling $\partial/ \partial x^{\mu} \rightarrow ({\partial_+}/{\partial x^{\mu}},{\partial_-}/{\partial x^{\mu}})$, which finally leads to the two-valuedness of the wavefunction itself $\psi \rightarrow (\psi_1,\psi_2)$, characterizing a (Pauli) spinor. 

\begin{figure}[!ht]
\begin{center}
\includegraphics[width=12cm]{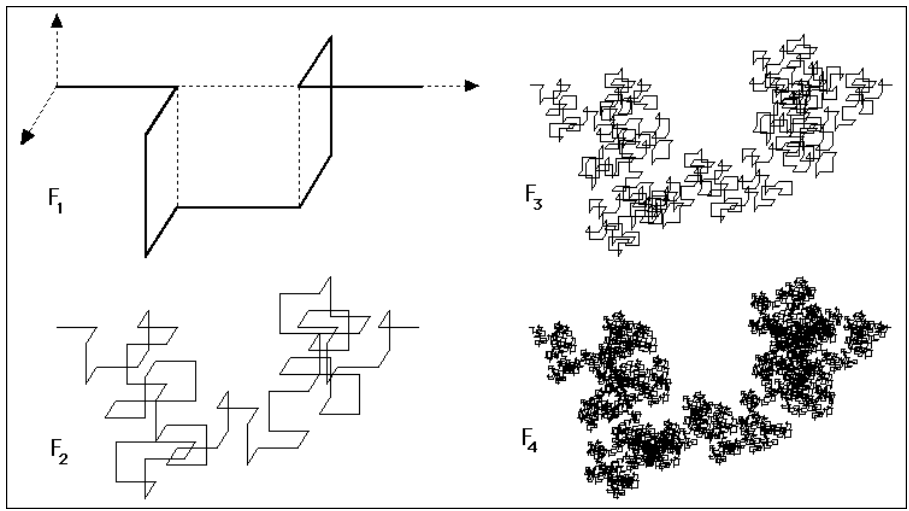}
\caption{\footnotesize{First four iterations on one period of an early model of infinitely spiral fractal curve (`hyperhelices'), from ~\cite{LN89}. Its generator is made of nine segments of length 1/3 and its fractal dimension is $D_F=2$. The spin of such a curve, whose fundamental period is a de Broglie wavelength $\lambda_{dB}=2 \pi\hbar/mv$, is in that case $\sigma=0.42 \; \hbar$. }}
\label{Paulifig1}
\end{center}
\end{figure}

A model for the emergence of a spin-like internal angular momentum (that was, however, not yet quantized in units of $\hbar/2$) in fractal spiral curves of fractal dimension 2 has been proposed in the 1980s \cite{LN89,LN93}. Note that this kind of fractal spiral curve has recently known a renewal of interest under the name `hyperhelices' \cite{NF04,CT06}. Let us briefly recall here the argument (see figure~\ref{Paulifig1}). The angular momentum $L_z=m r^2\dot{ \varphi}$ should classically vanish for $r\rightarrow 0$. But in the fractal spiral model, when a scale factor $q^{-1}$ is applied to the radius $r$, the number of turns and therefore the rotation velocity is multiplied by a factor $p=q^{D_F}$, so that  the angular momentum becomes multiplied by a factor $p\times q^{-2}=q^{D_F-2}$. It therefore remains defined at the infinitely small limit $q^{-1}\rightarrow 0$ in the special case $D_F=2$. In other words,  $\dot{ \varphi}\rightarrow \infty$ when  $r\rightarrow 0$ in such a way that the product $ r^2\dot{ \varphi}= 0^2 \times \infty$ remains finite when $D_F=2$ (while it is vanishing for $D_F<2$ and divergent for $D_F>2$). 

This model of spin therefore uses in an essential way the scale dependence of fractal geometry, that allows one to deal in a new manner with vanishing and infinite quantities (in particular, by showing that a general description of fractal geodesics actually leads to define non-differentiable wavefunctions which are still solutions of the standard equations of quantum mechanics \cite{LN05}). While in the standard differentiable approach the encounter of a zero or infinite quantity usually leads one to stop a calculation, the explicitly scale-dependent tools of the scale relativity theory allow one to go beyond such an obstacle and to prove the existence of finite and measurable quantities of the  ($0 \times \infty$) type. 

Note that the second-order terms in the quantum covariant total derivative (which are, in this framework, the basis of the Heisenberg relations) have exactly the same nature, namely $d\xi^2/dt$ is a differential element of first order in the differentiable theory, so that it should classically vanish. But in the fractal dimension 2 case, $d\xi^2/dt= (d\xi/dt)^2 \times dt = (2{\cal D}/dt) \times dt=2{\cal D}$ is now finite since the fractal velocity $d\xi/dt$ is now formally infinite (at the limit $dt \rightarrow 0$).

This result solves the problem of the apparent impossibility to define a spin in a geometric way both for an extended object and for a point-like object and provides another proof of the critical character of the value $D_F=2$ for the fractal dimension of quantum particle paths \cite{FH65}. 

It is also remarkable that the existence of spiral structures at all scales is also one of the elements of description of spinors in the framework of Ord's reformulation of the Feynman relativistic chessboard model in terms of spiral paths \cite{G092}.

\subsection{Numerical simulations of exact solutions}

We can now go beyond these fractal models of the spin and give a geometric physical picture of its nature based on explicit solutions of the Pauli or Dirac equation, since the fractal velocity fields of the non-differentiable spacetime geodesics can be derived from these solutions. It is remarkable that this now exact geometric description (whose spin is quantized in units of $\hbar/2$) supports the main features of the previous rough fractal models.

In order to exhibit this picture, we shall perform numerical simulations of the stochastic differential equations that have been set at the origin of the description. Recall that we have decomposed the elementary displacements in a fractal spacetime in terms of a classical (differentiable) part and of a fractal (non-differentiable) part,
\begin{equation}
dX_{\pm \pm}=v_{\pm \pm} \, dt + d \xi_{\pm \pm},
\end{equation}
where the geometric fractal fluctuation is replaced by a stochastic variable such that $\langle d\xi^2 \rangle/c dt= \lambda_c $ and $\langle d\xi \rangle=0$. Then the velocity fields $v_{\pm \pm}$, after they have been recombined as a unique biquaternionic velocity field, are solution of a bi-spinorial geodesics equation $\dfr {\cal V}/ds=0$ which can be integrated in terms of the Dirac equation, whose non-relativistic limit is finally the Pauli equation. Therefore, solving the Pauli equation for a given physical problem yields a quaternionic wavefunction
\begin{equation}
\psi = \phi_0 + i \chi_0 + (\phi_1 + i \chi_1)e_1 ,
\end{equation}
from which the velocity fields can be derived, for example,
\begin{align}
v^{\mu}_{++} = - {\frac{{\cal S}_0}{mc}} (
\phi_0 \, \partial^\mu\chi_0 + \chi_0 \, \partial^\mu\phi_0 + \phi_1 \, \partial^\mu\chi_1 + \chi_1 \, \partial^\mu\phi_1\nonumber\\ + 
\phi_0 \, \partial^\mu\phi_0 - \chi_0 \, \partial^\mu\chi_0 + \phi_1 \, \partial^\mu\phi_1 - \chi_1 \, \partial^\mu\chi_1) .
\label{eq.1065}
\end{align}
Then one can finally plot various realizations of the geodesics by performing numerical integrations of the stochastic differential equation 
\begin{equation}
dX_{++}=v_{++} \, dt + \eta \sqrt{\lambda_c c dt},
\label{stocdifeq}
\end{equation}
in which the explicit form of $v_{++}$, given by equation~(\ref{eq.1065}), is inserted ($\eta$ is a normalized stochastic variable such that $\langle \eta^2 \rangle=1$ and $\langle \eta \rangle=0$).

\begin{figure}[!ht]
\begin{center}
\includegraphics[width=7cm]{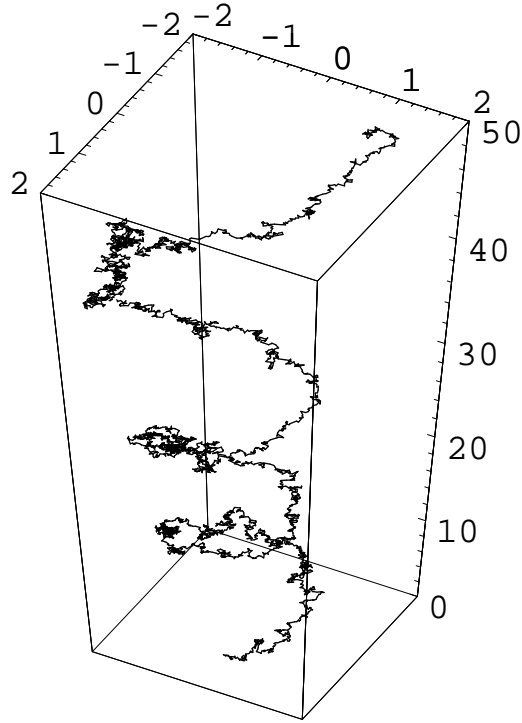}
\caption{\footnotesize{Numerical simulation of a typical spinorial geodesics in a fractal space. This is one realization among infinite possible realizations of the solutions of the stochastic differential equation (\ref{stocdifeq}). The plotted curve corresponds to the following values of the parameters: ${\cal D}=\hbar/2=0.05$, $dt=0.01$, $\sigma=1/2$, $v=1$. We have chosen a large value for the spin, $\sigma=5 \, \hbar$, in order to render the spiral shape fully apparent. Note that a `same' given curve may be plotted at infinite possible resolution values $dt$, and that there are infinite such fractal geodesics. }}
\label{Paulifig2}
\end{center}
\end{figure}

A general form of a spinor wavefunction has been given by Cohen-Tannoudji {\it et al} \cite{CDL73}, namely for a spin 1/2 particle,
\begin{equation}
|\psi\rangle = \cos({\theta}/{2})\; e^{-i \phi/2} \; |+\rangle + \sin({\theta}/{2}) \; e^{i \phi/2}  \; |-\rangle.
\end{equation}
A simplified case has been studied by Dezael \cite{FXD03}, who has considered the spinor
\begin{equation}
\psi= A_{0} \;exp\left(-\frac{i}{\hbar}(\overrightarrow{p_{0}}.\overrightarrow{r}-E_{0}t+\sigma_{0}\phi) \right)
+A_{1} \; exp\left(-\frac{i}{\hbar}(\overrightarrow{p_{1}}.\overrightarrow{r}-E_{1}t+\sigma_{1}\phi)\right).
 \end{equation}
From this expression, the biquaternionic velocity given by $ m \overrightarrow{V} \simeq
i\hbar\psi^{-1}\overrightarrow{\nabla}\psi$ is derived, which must be such that $\overrightarrow{v}_{-+} = 
-\overrightarrow{\tilde{v}}_{++}$ in the non-relativistic approximation considered here.  This is only possible provided 
$\overrightarrow{p_{0}}=\overrightarrow{p_{1}}$, and 
$\sigma_{0}=\sigma_{1}$ \cite{FXD03}. The bi-quaternionic velocity becomes
\begin{equation}
\overrightarrow{V} \simeq \frac{1}{m} 
\l(\overrightarrow{p_{0}}+\frac{\sigma_{0}}{r \sin\theta}\overrightarrow{u_{\phi}}\r)
\label{third dvpmt of V(phi)}
\end{equation}
where $\overrightarrow{u_{\phi}}$ is the unitary vector associated to the rotation by $\phi$ around $\overrightarrow{p_{0}}$. Dezael finally obtains
\begin{equation}
\overrightarrow{v}_{++}  \simeq 
\frac{1}{m} 
\l(\overrightarrow{p_{0}}+\frac{\sigma_{0}}{r \sin\theta}\overrightarrow{u_{\phi}}\r),
\label{result for v++}
\end{equation}
i.e., in Cartesian coordinates,
\begin{equation}
\left \{ \begin{array}{c}
\dot{x} = -\frac{\sigma_{0}}{m} \frac{y}{x^{2}+y^{2}} \\
\\
\dot{y} = \frac{\sigma_{0}}{m} \frac{x}{x^{2}+y^{2}} \\
\\
\dot{z} = \frac{p_{0}}{m} .
\end{array} \right.
\label{cartesian differential equation 2}
\end{equation}
These equations clearly describe a class of spiral motions such that $m r^2 \dot{\phi}=\sigma_0$, where $r$ can take any value, so that $r\rightarrow 0$ implies $\dot{\phi}\rightarrow \infty$, in agreement with the early spiral models \cite{LN89}.

Then we finally carry over this expression for the classical velocity field into the stochastic differential equation (\ref{stocdifeq}) and integrate it numerically. A typical example of the spiral fractal paths obtained in this process is shown in figure~\ref{Paulifig2}. 


\section{Discussion and conclusion}
\label{s.dac}

One would require a genuine fundamental physical theory, not only to be able to derive from physical principles the correct equations of physics, but also that some physical mechanism would always prevent an equation to be written when it is unphysical. Such a requirement is clearly impossible when the physical foundation of a physical theory remains axiomatic. This is the case of standard quantum mechanics, where no physical principle prevents {\it a priori} from writing a Pauli equation with the wrong magnetic moment. It is the experiment which proves such an equation to be wrong, and also the experiment which proves the Dirac equation and its non-relativistic limit to be correct.

The attempts of the scale relativity theory to found quantum mechanical laws on first principles allow one to come back on this question. We have shown in this paper that it was indeed impossible to directly write a non-relativistic equation for spin-1/2 particles, and that it could therefore only be derived as a non-relativistic limit of the relativistic equation.

In the framework considered here, the non-relativistic case corresponds to considering only a fractal three-dimensional space, without yet introducing fractality for the time variable (which is identified in this case of Galilean approximation with an invariant proper time), while the relativistic case is identified with working in a full four-dimensional spacetime. The reason for this identification is simply that the main transition from the (small scale) fractal to (large scale) non-fractal and classical regimes occurs around the Einstein-de Broglie scale \cite{LN89,LN93}, i.e. $\hbar/p$ for space and $\hbar/E$ for time. It is therefore the very existence of mass, through the relation $m^2=E^2-p^2$(for $c=1$), which leads to a fundamental disymmetry between space and time as concerns the scale space. When going from large scales to small scales, one first encounters the space transition (as exemplified e.g. by the fact that atomic physics is mainly non-relativistic), then at scales smaller than the Compton scale $\hbar/m$ (which is the Einstein scale in rest frame) the time transition to full fractal spacetime and relativistic quantum mechanics.

Now, in the scale relativity theory, the inclusion of spin in the description comes from the account of the differential parity (mirror) discrete symmetry breaking of the transformation $dx\rightarrow -dx$ which is a direct consequence of the non-differentiable geometry. It is a generalisation of the equivalent (proper) time symmetry breaking of the transformation $ds \rightarrow -ds$, that gives birth to the complex nature of the wavefunction \cite{LN93,CN04}.

Accounting for all these effects in fractal spacetime has led to the construction of bi-spinors (described as complex quaternions) and to the derivation of the Dirac equation \cite{CN04}. Taking its non-relativistic limit (which we have explicitly done in the present paper) yields Pauli spinors and the Pauli equation with the correct magnetic moment including the relativistic factor 2. Now the question addressed in this paper was whether the account of both the space and time discrete symmetry breakings on the differential elements was possible in a fractal space (without including fractality on time). The answer is clearly no, since the integration of the basic geodesics equation in terms of a Schr\"odinger form has revealed to be impossible in that case (which also means the impossibility to define a function $\psi$ having all the properties of a wavefunction). 

Following the requirement of coherence of fundamental theories recalled at the beginning of this conclusion, we interpret this apparently negative result as a success of the theory, and as a direct proof of the fundamentally relativistic nature of the spin, which remains true and relevant even in a non-relativistic situation.

In section \ref{s.gfs}, we have given an example of numerical integration of a typical fractal geodesics (among infinite possible geodesics whose set builts the wavefunction) in the spinorial case, which is characterized by multiscale fractal spiral patterns. This allows one to have a physical geometric picture of what spin is. Such a picture may help understanding more thoroughly the various features of EPR spin experiments (which, as already remarked in ~\cite{CN04}, is accounted for in its essence in the scale relativity framework since it alows one to derive the standard properties of Dirac bi-spinors). This will be developed in a forthcoming work, including a scale-relativistic analysis of density matrices, which are at the heart of some of the subtleties of the EPR experiment. We may also contemplate the possibility of using such a spin representation to achieve future experimental tests of the fractal approach. Indeed, even though the true spin nature of elementary particles is probably not reproducible since it involves the full infinities of the non-differentiable geometry (without any lower limit), one could consider an experimental (artificial) construction of spiral trajectories which would be fractal on a large enough range of internal scales so that some of the properties of a quantum spin could be recovered as an approximation.



\end{document}